\documentclass[10pt,preprint2]{aastex}
\shorttitle{Low-amplitude Variables in NGC 6791}
\shortauthors{Mochejska et al.}
\begin{document}

\title{Planets in Stellar Clusters Extensive Search. I.~Discovery 
of 47 Low-amplitude Variables in the Metal-rich Cluster
NGC~6791 with Millimagnitude Image Subtraction Photometry.}

\author{B. J. Mochejska}
\affil{Copernicus Astronomical Center, Bartycka 18, 00-716 Warszawa}
\email{mochejsk@camk.edu.pl}
\author{K. Z. Stanek, D. D. Sasselov\altaffilmark{1} 
\& A. H. Szentgyorgyi}
\affil{Harvard-Smithsonian Center for Astrophysics, 60 Garden St.,
Cambridge, MA~02138}
\email{kstanek@cfa.harvard.edu, sasselov@cfa.harvard.edu,
saint@cfa.harvard.edu}
\altaffiltext{1}{Alfred P. Sloan Research Fellow.}

\begin{abstract}
We have undertaken a long-term project, Planets in Stellar Clusters
Extensive Search (PISCES), to search for transiting planets in open
clusters. As our first target we have chosen NGC 6791 -- a very old,
populous, metal rich cluster. In this paper we present the results of
a test observing run at the FLWO 1.2 m telescope. Our primary goal is
to demonstrate the feasibility of obtaining the accuracy required for
planetary transit detection using image subtraction photometry on data
collected with a 1 m class telescope. We present a catalog of 62
variable stars, 47 of them newly discovered, most with low amplitude
variability. Among those there are several BY Dra type variables. We
have also observed outbursts in the cataclysmic variables B7 and B8
(Kaluzny et al. 1997).
\end{abstract}

\hyphenation{XVIIth}
\section{{\sc Introduction}}
Since antiquity mankind has wondered whether planetary systems other
than our own exist. The first documented effort aimed at extrasolar
planet detection was undertaken by Huygens in the XVIIth
century. Starting in the 1930s, subsequent searches have been
attempted, but failed to produce positive results due to insufficient
measurement precision.  Only recently it has become possible to obtain
radial velocity measurements accurate enough to indirectly detect
planets via Doppler shifts of stellar spectra of stars other than the
Sun (Mayor \& Queloz 1995).

To date, over 70 planets have been discovered, mainly around solar
type stars. All of them were detected by radial velocity surveys (eg.
Marcy et al.~2001, Noyes et al.~1997). For one of these systems, HD
209458, the transit of the planet across the host star's disk has been
observed (Charbonneau et al.~2000, Henry et al. 2000), thus
demonstrating the feasibility of detecting planets this way. Several
groups are currently monitoring the brightness of thousands of stars
to search for planets via transits (eg. Brown \& Charbonneau 1999,
Quirrenbach et al.~1998). Recently, Udalski et al.\ (2002) discovered
46 stars with transiting low-luminosity companions. These objects may
be planets, brown dwarfs or M dwarfs. A confirmation of their nature
will be provided by mass determinations based on photometry combined
with radial velocities from followup spectroscopic observations.

The analysis of the properties of stars with planets suggests that
they are on the average significantly more metal rich than those
without (Santos et al.~2001). Some studies indicate that the source of
the metallicity is most likely ``primordial'' (Santos et al.~2001,
Pinsonneault et al.~2001), while others suggest that the observed high
metallicity is intrinsic only in some cases, with the more likely
cause being the accretion of planetesimals onto the star (Murray \&
Chaboyer 2001) or the infall of giant gas planets (Lin 1997).

Although the observed lack of planets in the low metallicity
([Fe/H]$=-0.7$) globular cluster 47 Tuc (Gilliland et al.~2000) is
compatible with the ``primordial'' metallicity scenario, the case is
far from being resolved. In such dense environments as the cores of
globular clusters encounters with other stars may lead to the breakup
of planetary systems (Davies \& Sigurdsson 2001). The study of open
clusters offers the possibility of disentangling the effects of
metallicity and crowding: their stellar densities are not high enough
for crowding-induced disruption to be effective.

We have undertaken a long-term project, Planets in Stellar Clusters
Extensive Search (PISCES), to search for transiting planets in open
clusters. As our first target we have chosen NGC 6791
$[(\alpha,\delta)_{2000}=(19^h20.8^m, +37^{\circ}51')]$, a very old,
extremely metal rich cluster ($\tau$=8 Gyr, [Fe/H]=+0.4; Chaboyer et
al.~1999). At a distance modulus of (m-M)$_V$ = 13.42 (Chaboyer et
al.~1999) it contains about 10000 stars (Kaluzny \& Udalski 1992,
hereafter KU92).

In this paper we present the results of a test observing run at the
FLWO 1.2 m telescope. Our aim was to demonstrate the feasibility of
obtaining the required accuracy using image subtraction photometry on
data collected with a 1-m class telescope needed to reliably detect
transits of inner-orbit gas-giant planets with an acceptably low false
alarm rate.

On clear nights we have reached the desired level of photometric
precision. Unfortunately, there were few such nights during our 26
night run, which is typical for July at FLWO. Even though this dataset
is not optimal for planetary transit detection, it allowed us to
discover 47 new low amplitude variables, compared to 22
previously known (Kaluzny \& Rucinski 1993; Rucinski, Kaluzny \&
Hilditch 1996, hereafter: RKH).

The paper is arranged as follows: \S 2 describes the
observations, \S 3 summarizes the reduction procedure, \S 4
outlines the procedure of variable star selection, \S 5 contains
the variable star catalog. Concluding remarks are found in \S 6.

\section{{\sc Observations}} 
\begin{figure}[t]
\plotone{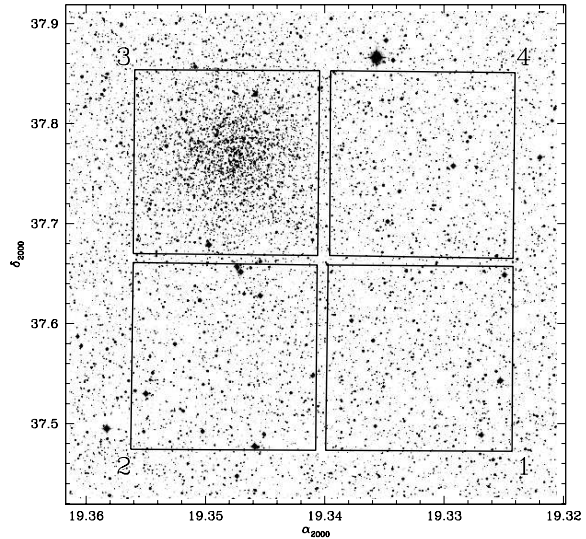}
\caption{Digital Sky Survey image of NGC 6791 showing the field of
view of the 4Shooter. The chips are numbered clockwise from 1 to 4
starting from the bottom left chip. NGC 6791 is centered on chip 3.
North is up and east is to the left}
\label{chips}
\end{figure}

The data analyzed in this paper were obtained at the FLWO 1.2 m
telescope using the 4Shooter CCD mosaic with four thinned, back side
illuminated AR coated Loral $2048^2$ CCDs (Szentgyorgyi et al.~2002).
The camera, with a pixel scale of $0\farcs 333$ pixel$^{-1}$, gives a
field of view of $11\farcm 4\times 11\farcm 4$ for each chip. The
cluster was centered on chip 3 (Fig. \ref{chips}). The data were
collected during 18 nights, from July 6th to August 1st, 2001. None of
the nights were photometric and most were at least partially cloudy, as
this was the monsoon season. A total of $204\times 900$ s $R$ and
$36\times 450$ s $V$-band exposures were obtained. The median seeing was
$1\farcs 9$ in R and $2\farcs 1$ in V.

\section{{\sc Data Reduction}}

\subsection{{\it Image Subtraction Photometry}}
The preliminary processing of the CCD frames was performed with the
standard routines in the IRAF ccdproc package.\footnote{IRAF is
distributed by the National Optical Astronomy Observatories, which are
operated by the Association of Universities for Research in Astronomy,
Inc., under cooperative agreement with the NSF.}

The photometry for all stars in the field was extracted using the ISIS
image subtraction package (Alard \& Lupton 1998, Alard 2000) from the
$R$ and $V$-band data for all four CCD chips. 

ISIS is based on the fast optimal image subtraction (OIS)
algorithm. In order to successfully subtract two images, it is
necessary to exactly match their seeing. In OIS this is accomplished
by finding a convolution kernel $Ker$ and difference in background
levels $bg$ which will minimize the squared differences between both
sides of the equation:
\begin{eqnarray*}
Im(x,y) = Ker(x,y;u,v)\otimes Ref(u,v) + bg(x,y)
\end{eqnarray*}
where $Ref$ is the reference image and $Im$ is the processed image.
(Alard \& Lupton 1998, Wozniak 2000).

The ISIS reduction procedure consists of the following steps: (1)
transformation of all frames to a common $(x,y)$ coordinate grid; (2)
construction of a reference image from several best exposures; (3)
subtraction of each frame from the reference image; (4) selection of
stars to be photometered; (5) extraction of profile photometry from
the subtracted images.

All computations were performed with the frames internally subdivided
into four sections ($sub\_x=sub\_y=2$). Differential brightness
variations of the background were fit with a first degree polynomial
($deg\_bg=1$). A convolution kernel varying quadratically with
position was used ($deg\_spatial=2$).

An image of particularly good quality was selected as the template
frame for the stellar positions. The remaining images were re-mapped
to the template frame coordinate system using a second degree
polynomial transform ({\sc degree}=2). During this step an initial rejection
of cosmic rays was also performed. A setting of 1.0 for the cosmic ray
threshold ({\sc cosmic\_thresh}) was used.

A reference image was then constructed from 20 best images in $R$ and
15 in $V$. The constituent images were processed to match the template
PSF and background level and then stacked by taking a median in each
pixel to obtain a reference image virtually free of cosmic rays.

Image subtraction was then applied to all the frames. For each frame
the reference image was convolved with a kernel to match its PSF and
then the frame was subtracted from it. As the flux of the non-variable
stars on both images should be almost identical, such objects will
disappear from the subtracted image. The only remaining signal will
come from variable stars.

Next comes the selection of stars for photometry. Normally at this
stage the standard ISIS variable detection procedure is used. It
computes a median of absolute deviations on all subtracted images and
performs a simple rejection of cosmic rays and defects.  This approach
works very well for variables with an amplitude of at least 0.1 mag,
but is much less efficient for variables of smaller amplitude. As we
are interested in these latter variables we extracted photometry for
all the stars on the template list, to search them for variability
using more traditional methods.

\begin{figure*}[t]
\epsscale{2}
\plottwo{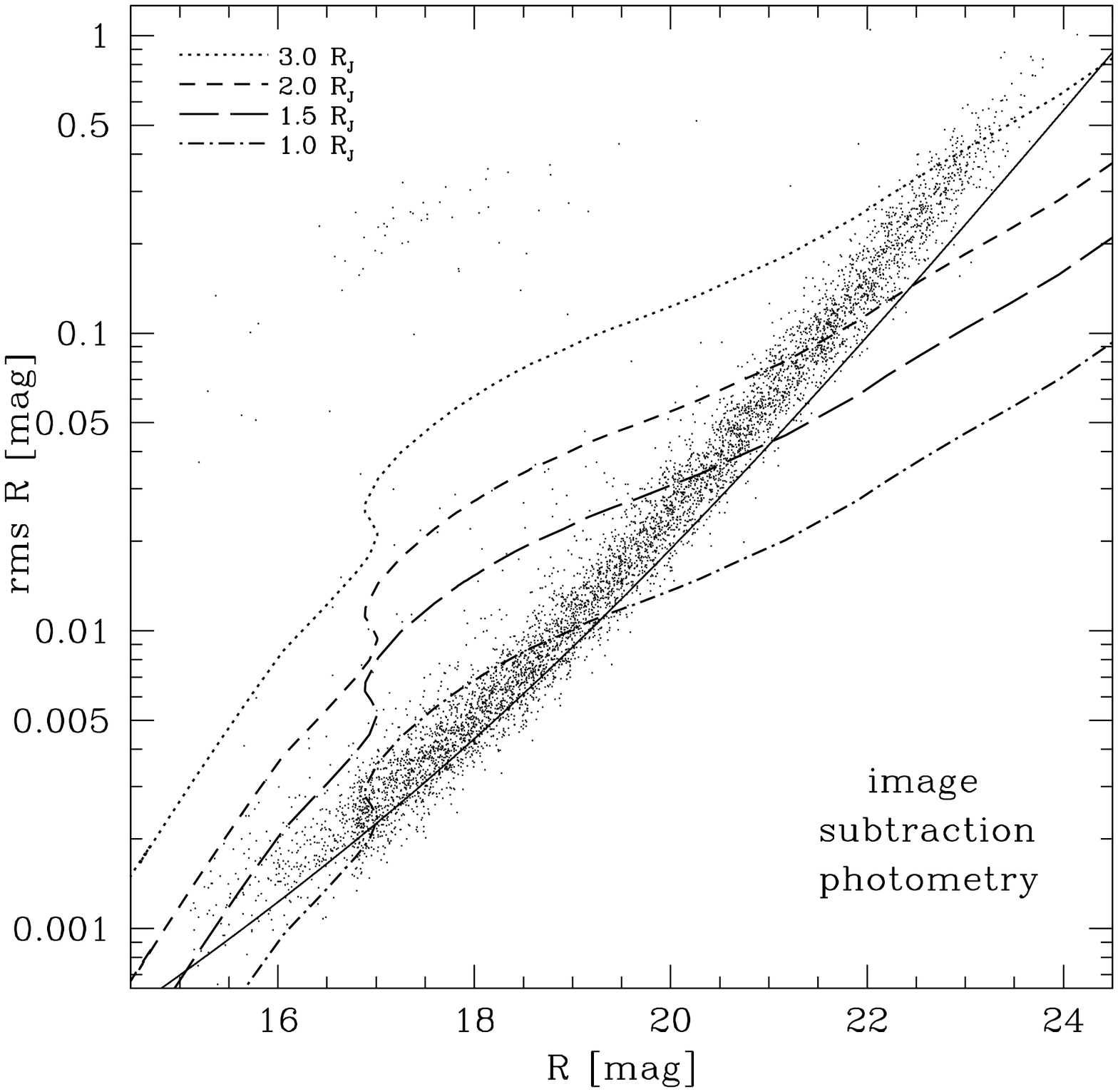}{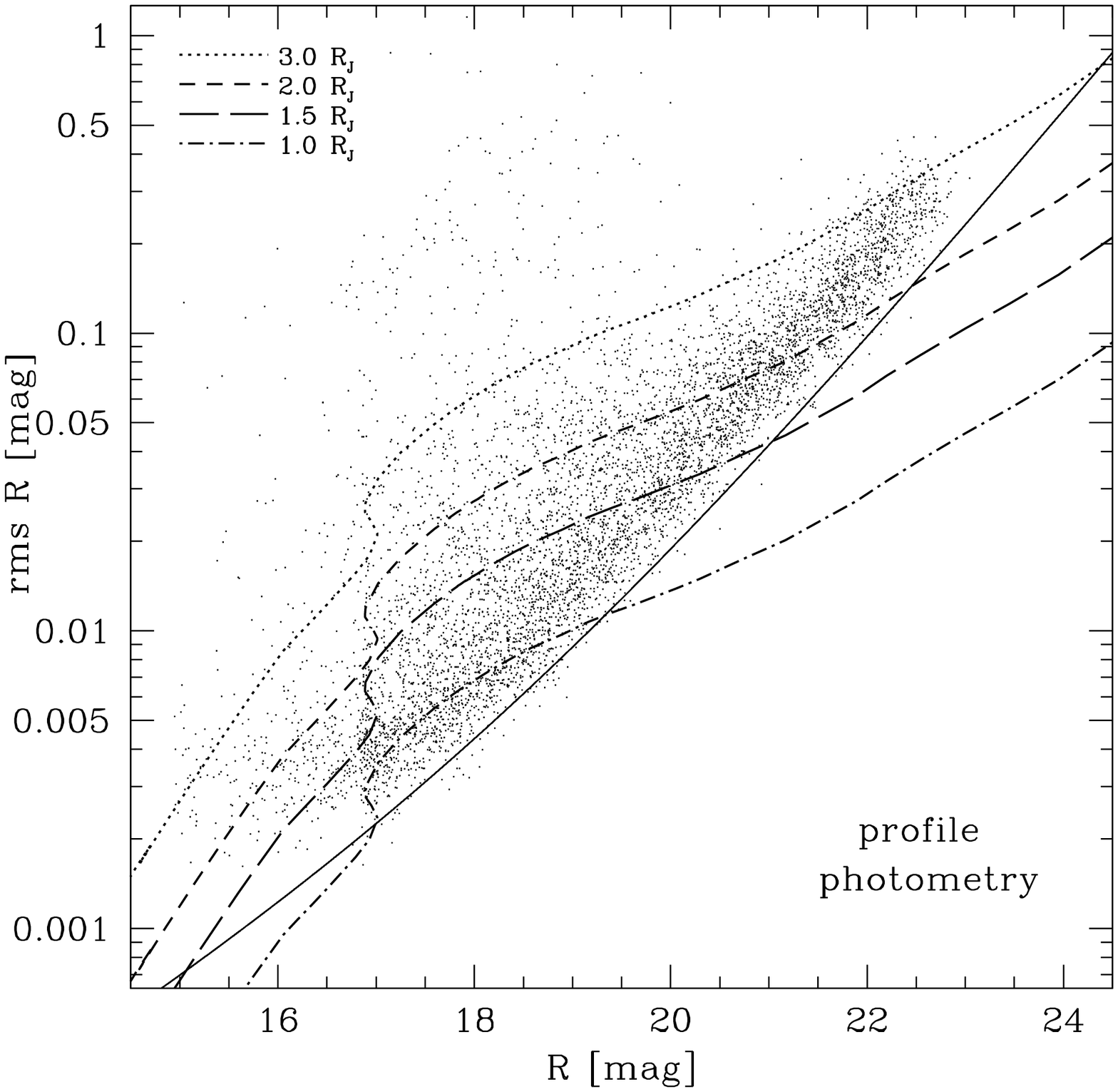}
\epsscale{1}
\caption{The rms scatter of the $R$-band light curves for the best
night. The left panel shows the scatter for ISIS light curves, the
right panel for {\sc DAOphot}. The continuous curve indicates the
photometric precision limit due to Poisson noise of the star and sky
brightness and the dashed ones show the 6.5 $sigma$ detection limits
for 1, 1.5, 2 and 3 $R_J$ planets with orbital periods of 3.5 days.
}
\label{rms}
\end{figure*}

A final step, not included in the ISIS image subtraction package
itself, was to convert the light curves from ADU to magnitudes.  For
this purpose the template frames were reduced with the {\sc
DAOphot/Allstar} package (see Section 3.2 for details).  The template
instrumental magnitude $m_{tpl}$ of each star was converted into
counts $c_{tpl}$, using the {\sc Allstar} zeropoint of 25 mag. The
light curve was then converted point by point to magnitudes $m_i$ by
computing the total flux $c_i$ for a given epoch as the sum of the
counts on the template $c_{tpl}$ and the counts on the subtracted
template image $\Delta c_{tpl} = c_{ref}-c_{tpl}$ decreased by the
counts corresponding to the subtracted image $\Delta c_i=c_{ref}-c_i$:
\begin{equation}
c_i = c_{tpl} + \Delta c_{tpl} - \Delta c_i
\end{equation}
The flux was converted to instrumental magnitudes using the same zero
point as above. To convert the photometric error $\sigma^c_i$
expressed in counts to $\sigma^m_i$ in magnitudes, we used the
following relation:
\begin{equation}
\sigma^m_i = -2.5 \log \left( \frac{c_i}{c_i+\sigma^c_i}\right) 
\end{equation}

\subsection{{\it Profile Photometry}}
Profile photometry was extracted using the {\sc DAOphot/Allstar}
package (Stetson 1987) from images remapped onto the template $(x,y)$
coordinate grid. This was done mainly to compare its precision to that
of image subtraction photometry.

A PSF varying linearly with position on the frame was used. Stars were
identified using the FIND subroutine and aperture photometry was
performed with the PHOT subroutine. The same star list was used on all
frames for the construction of the PSF. Of those the stars with
profile errors greater than twice the average were rejected and the
PSF was recomputed. This procedure was repeated until no such stars
were left on the list. The PSF was then further refined on frames with
all but the PSF stars subtracted from them. This procedure was applied
twice. The resultant PSF was then used by {\sc Allstar} in profile
photometry.

The next step was to obtain a template list of stars, to be also used
as the input star list for the photometry routine in ISIS. The
template and the stacked reference image were reduced following the
same procedure as for single frames. The stars were then subtracted
from the reference image and an additional 100-200 stars undetected in
the initial FIND run were hand-selected and added to the star
list. {\sc Allstar} was run again twice on the expanded list, which
resulted in the rejection of some more objects. The star list was
further cleaned by rejecting objects with $\chi>3$ and with errors too
large for their magnitude. This list was used as input to {\sc
Allstar} on the template frame to create the final list of stars.

The final template star list was then used as input to {\sc Allstar}
in the fixed-position mode on each of the frames. The output profile
photometry was transformed to the common instrumental system of the
template image and then combined into a database. The database was
created for the $R$-band images in chip 3 only.

\begin{figure}[th]
\epsscale{1}
\plotone{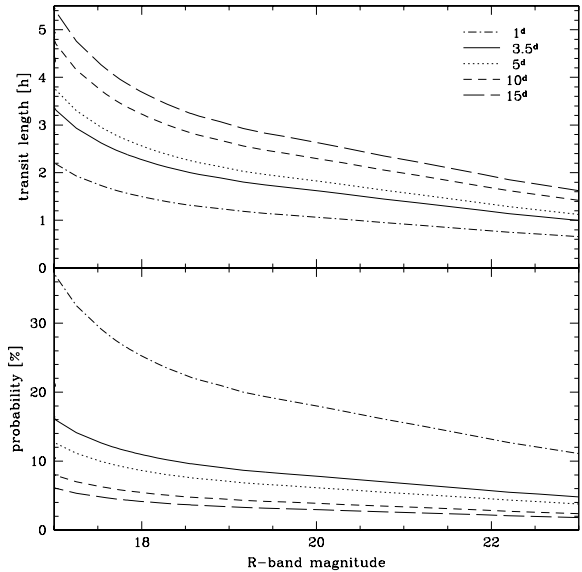}
\caption{ The transit length (upper panel) and the probability of 
occurence of a transit given random inclinations (lower panel) as a
function of $R$-band magnitude, plotted for periods of 1, 3.5, 5, 10
and 15 days.}
\label{fig:tau}
\end{figure}

\begin{figure}[th]
\epsscale{1}
\plotone{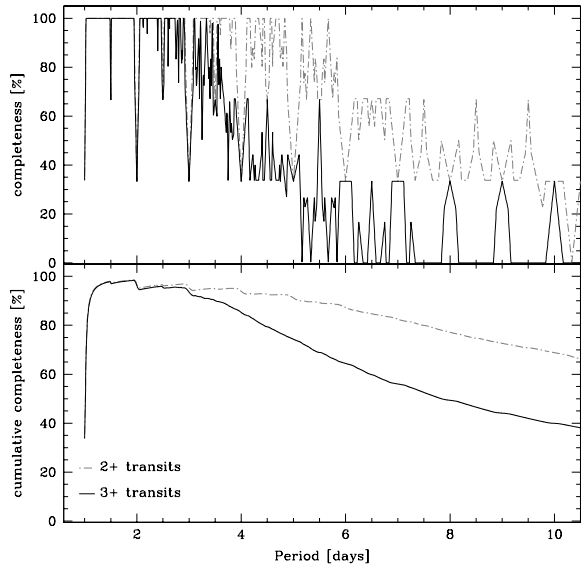}
\caption{Detection efficiency of planetary transits as a function of their
period, assuming $8^h$ of observations per night during 30 consecutive
nights.}
\label{fig:comp}
\end{figure}

\begin{figure}[th]
\epsscale{1}
\plotone{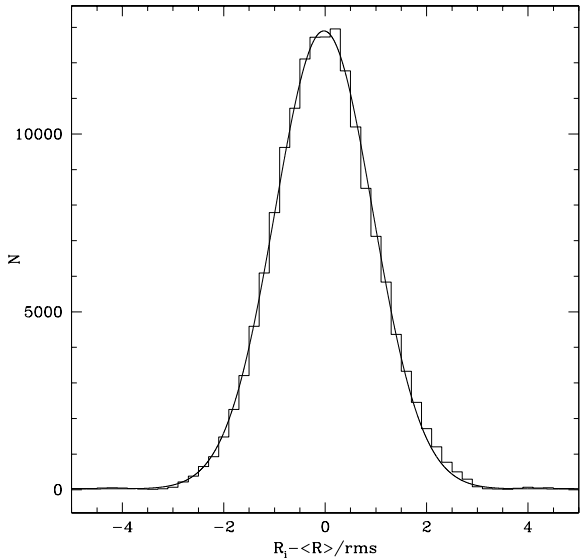}
\caption{ The distribution of best night residuals for 6871 stars
normalized to the rms. }
\label{fig:gauss}
\end{figure}

\begin{figure}[t]
\plotone{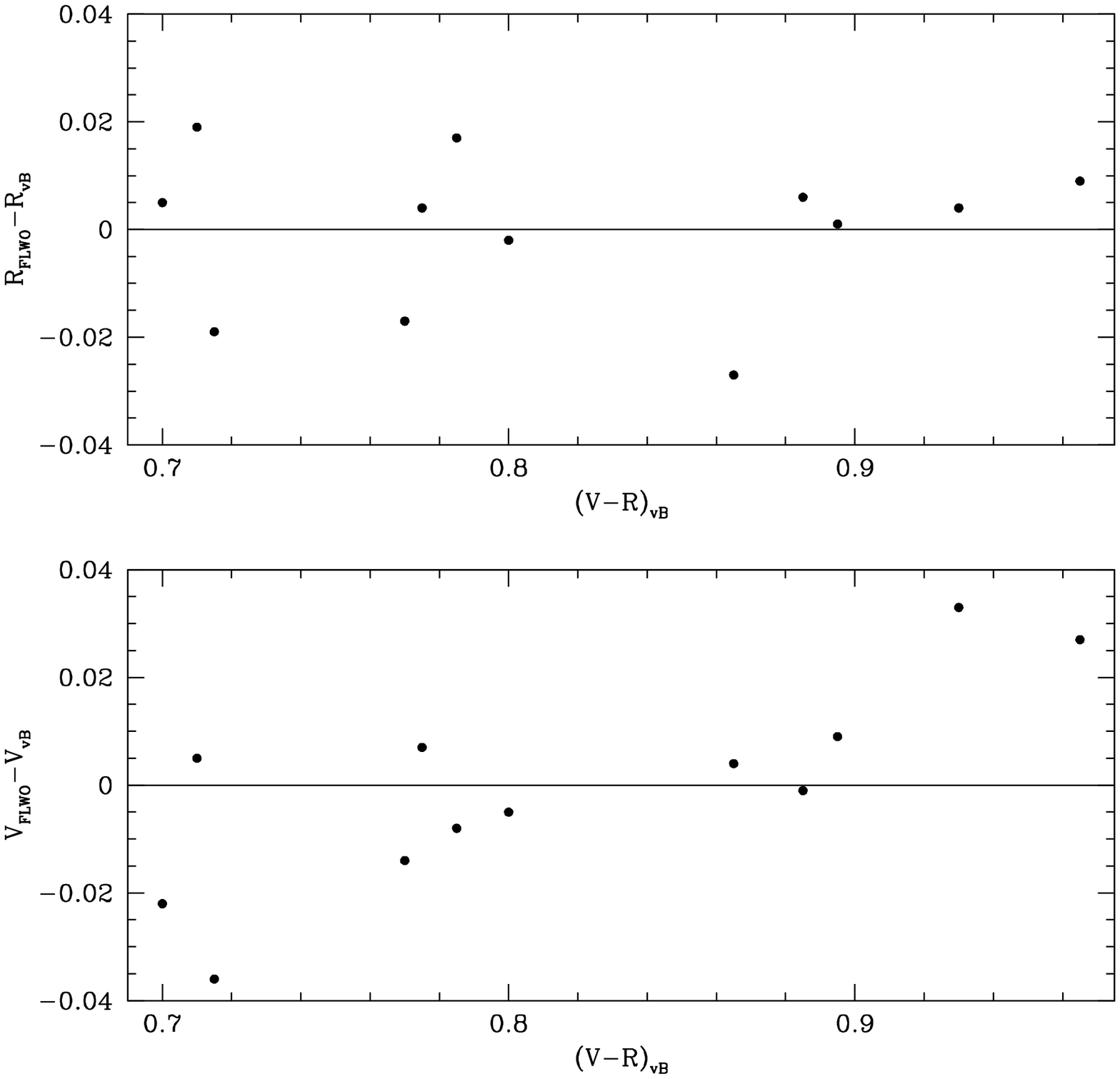}
\caption{The $R$ and $V$ residuals as a function of $\vr$ for 12 stars
from von Braun et al. (1998).}
\label{cal_rgb}
\end{figure}

%\subsection{{\it Comparison of Image Subtraction and Profile Photometry}}
\subsection{{\it Implications for transit detection}}

In Figure \ref{rms} we plot the ISIS (left panel) and {\sc
DAOphot/Allstar} (right panel) time series precision as a function of
magnitude for the best night. This night should be representative of
the average night we expect other than during the July-August monsoon
season. The time series precision was computed as the rms of the light
curve, with the rejection of $3\sigma$ outliers. Only stars with light
curves containing at least 22 of 23 data points are plotted. The
continuous curve indicates the photometric precision limit due to
Poisson noise of the star and sky brightness. The dashed curves
correspond to the rms value which would yield a 6.5 $\sigma$ detection
for three transits in the $R$-band, for planets with 1, 1.5, 2 and 3
$R_J$ and orbital period of 3.5 days, the same as for HD 209458. The
curves are defined by the equation
\begin{equation}
rms = \sqrt{N} \frac{\Delta R}{\sigma}
\end{equation}
where $N$ is the number of observations during three transits, $\Delta
R$ is the $R$-band amplitude of the transit and $\sigma$ is set at
6.5. The steep fall of detection capability at the bright end is
caused by the increase in radius on the subgiant branch and the 8 hour
limit on the transit length set by the length of the night. The
increase at the faint end is caused by the decrease of the stellar
radius along the main sequence.

Not surprisingly, the overall precision is better and the scatter
smaller for image subtraction photometry. This is especially well seen
for the brightest stars. In the plot for image subtraction photometry
6562 (95\%), 5304 (77\%), 4110 (59\%) and 2053 (29\%) stars fall
within the detection limit for transits of 3, 2, 1.5 and 1 $R_J$
planets, respectively. 

There is a group of stars with $rms > 0.1$ mag and $R < 20$, with
several discrepant points in the light curve caused by bad columns,
which were not removed by sigma clipping. Following the approach of
Udalski et al.\ (2002) we will set an upper limit on the $rms$ to
remove such stars from the sample searched for planetary transits.

Figure \ref{fig:tau} shows the transit length (upper panel) and the
probability of occurence of a transit given random inclinations (lower
panel) as a function of $R$-band magnitude, plotted for periods of 1,
3.5, 5, 10 and 15 days. The transit length is proportional to the
period $P$ as $P^{\frac{1}{3}}$ and the probability of observing a
transit given random inclinations, is proportional to
$P^{-\frac{2}{3}}$. Both of these quantities depend on the mass, $M$,
and radius, $R$, of the star as $M^{-\frac{1}{3}} R$ (Eqs (1) in
Gilliland et al.\ 2000). The duration of the transits is of the order
of 1 to 4 hours and it decreases with the magnitude of the star. For
periods over $3.5$ days the probability of a transit is of the order
of 10\% for stars near the turnoff and it falls to 2-5\% for lower
main sequence stars.

Figure \ref{fig:comp} shows the detection efficiency (upper panel) and
cumulative detection efficiency (lower panel) for transiting planets
with periods greater than $1^d$ as a function of period, assuming
$8^h$ of observations per night during a $30$ day continuous run. The
choice of the minimum period is motivated by the shortest period
transiting object ($P=0.8082^d$) reported by Udalski et al. (2002),
although it should be noted that the shortest period for a planet
detected by radial velocity searches is $\sim3$ days (HD 83443, Mayor
et al. 2000). The detection efficiency for at least two transits drops
below 30\% at $P\sim9^d$ (upper panel, dot-dashed line) and for at
least three transits at $P\sim5^d$ (solid line). Such temporal
coverage would enable us to observe three or more transits for 50\% of
all transiting planets with periods between $1$ and $7.8$ days (lower
panel, solid line). A 50\% efficiency for the detection of at least
two transits is reached at a period of $15.7$ days (dot-dashed line).

The distribution of best night residuals for 6871 stars, normalized to
the rms, is shown in Fig.~\ref{fig:gauss}. Only stars with at least 22
out of 23 data points were included in the histogram. The distribution
of the residuals is well-fit by a Gaussian distribution with
$\sigma=0.96$.

\begin{figure*}[ht]
\epsscale{1.5}
\plotone{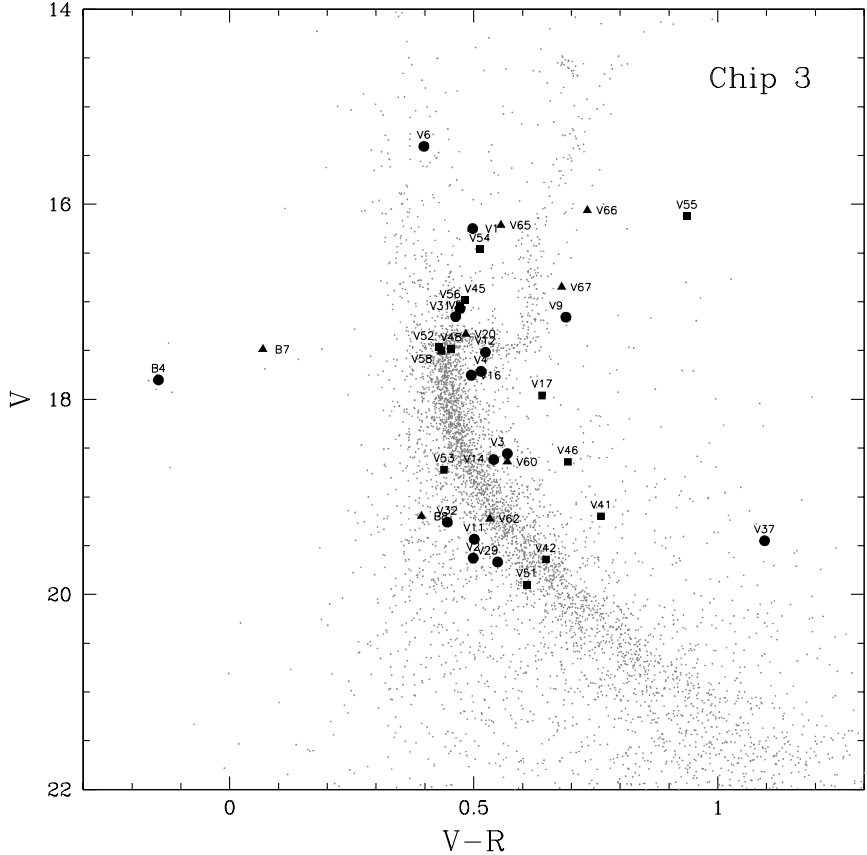}
\caption{$V/\vr$ color-magnitude diagram (CMD) for chip 3, centered on 
NGC 6791. Eclipsing binaries are plotted with circles, other periodic
variables with squares and miscellaneous variables with triangles.}
\label{cmd3}
\end{figure*}

\begin{figure*}[ht]
\epsscale{2}
\plotone{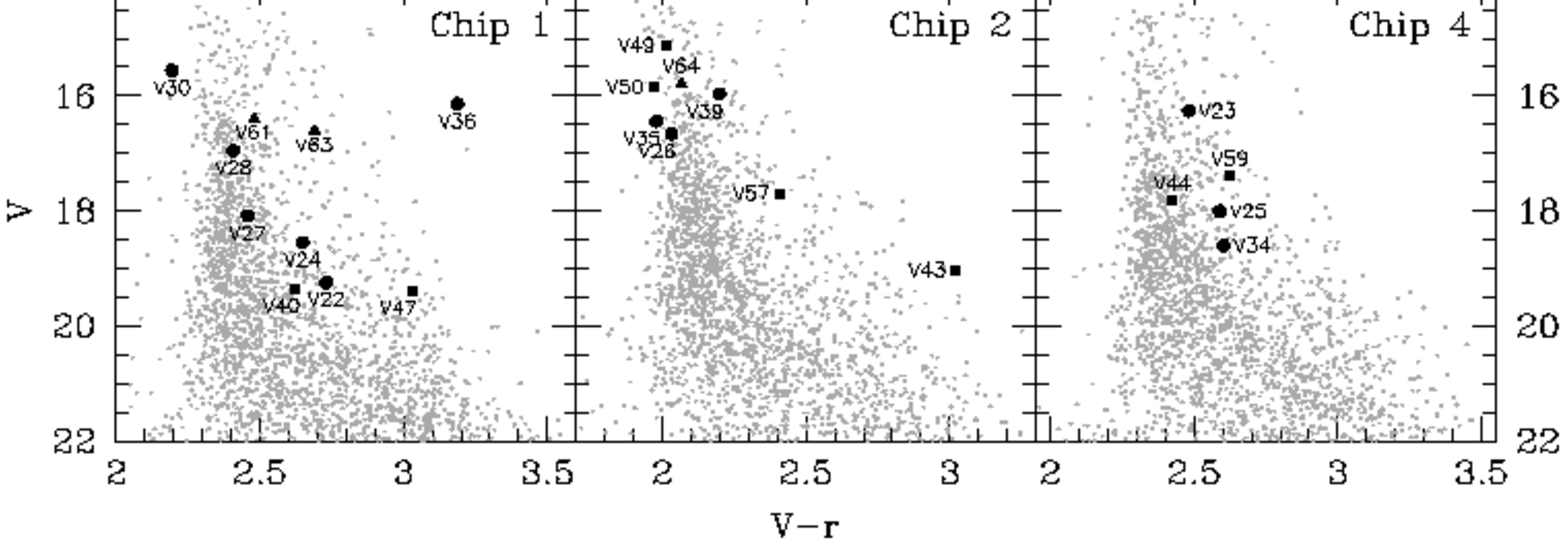}
\caption{$V/V-r$ CMDs for chips 1, 2 and 4. Eclipsing binaries are
plotted with circles, other periodic variables with squares and
miscellaneous variables with triangles.}
\label{cmd124}
\end{figure*}

\subsection{{\it Calibration}}
None of the nights in our observing run were photometric, thus we were
forced to rely on other, indirect sources of calibration.

The $R$-band photometry in chip 3 was calibrated from the photometry
for 12 of the 14 red giant branch (RGB) stars published by von Braun
et al.~(1998). One of the stars could not be identified on our
template and another one, much redder than the others (R4,
$\vr=1.355$) was rejected because it was a clear outlier in both $R$
and $V$. Due to the small number of stars and a very limited color
baseline, only an offset was determined.  The residuals, with an rms
scatter of 0.014 mag, are shown in the upper panel of
Fig. \ref{cal_rgb}. The $R$-band photometry for chips 1, 2 and 4 was
left uncalibrated, as they contain no stars with calibrated $R$-band
photometry.

A similar comparison was made for the $V$-band magnitudes of the 12
RGB stars and there seems to be a trend with color (lower panel of
Fig. \ref{cal_rgb}) and magnitude, as those two quantities are
correlated for RGB stars. A trend in magnitude is also present for
stars with $V<17$ in comparison with data of Kaluzny \& Rucinski
(1995; hereafter KR95), but is not seen in case of the KU92
photometry. Since the KR95 data cover parts of all four chips and the
offset is in good agreement with the von Braun et al.~(1998)
photometry (to $0.005$ mag), we have used offsets from the KR95
photometry to bring the instrumental magnitudes to the standard
system.

In Figures \ref{cmd3} and \ref{cmd124} we present the $V/\vr$ CMDs for
chip 3 and chips 1, 2, 4, respectively. The CMD for chip 3, centered
on the cluster, shows a well defined main sequence (MS) down to $V\sim
20$, a subgiant branch (SGB), a red giant branch and a red clump
at $V\sim 14.5$, $\vr\sim 0.7$. The blue stars, noted by KU92, are
also present at $\vr<0$. The CMDs for the remaining chips consist in
large proportion of disk stars; only an indication of the upper MS is
discernible.

\subsection{{\it Astrometry}}
Equatorial coordinates were determined for the $R$-band template star
lists, expanded with the variables with no $R$-band photometry. The
transformation from rectangular to equatorial coordinates was derived
using 734, 729, 1058 and 657 transformation stars from the USNO A-2
catalog (Monet et al. 1996) in chips 1 through 4, respectively. The
average difference between the catalog and the computed coordinates
for the transformation stars was $0\farcs 14$ in right ascension and
$0\farcs 12$ in declination.

\section{{\sc Variability Search}}
\subsection{{\it Selection of Variables}}
To select candidate variable stars we have employed the variability
index J defined by Stetson (1996). Below we present only a summary of
the method; the reader is referred to the original paper for details.

The variability index J is computed as follows:
\[J=\frac{\sum^n_{k=1}w_k \mathrm{sgn}(P_k)\sqrt{|P_k|}}{\sum^n_{k=1}w_k}\]
where the user has defined n pairs of observations, i and j, to be
considered, each with a weight $w_k$. In our case, observations
separated by less than 0.03 days were treated as a pair. A weight
$w_k=1$ was assigned to pairs of observations ($i(k)\neq j(k)$) and
$w_k=0.25$ to single observations ($i(k)=j(k)$). $P_k$ is the product
of the normalized residuals of the two observations, i and j, 
constituting the k-th pair:
\[P_k = \left\{ \begin{array}{ll}
\delta_{i(k)}\delta_{j(k)}, & \mathrm{if~} i(k)\neq j(k)\\
\delta_{i(k)}^2-1, & \mathrm{if~} i(k)=j(k)
\end{array} \right.\]
and $\delta$ is the magnitude residual of a given observation from the
average, scaled by the standard error:
\[ \delta = \sqrt{\frac{n}{n-1}}\frac{v-\bar{v}}{\sigma_v}\]
The final variability index was multiplied by a factor $\sum w/w_{max}$,
where $w_{max}$ is the total weight of a star if it were measured
on all images.

We are aware that this approach is not optimal for detecting planetary
transits and a better method based on the matched-filter algorithm is
under development. Recently, Udalski et al.\ (2002) used this
technique to identify 46 transiting planet candidates. They report
that their implementation was very sensitive even to single transit
events and produced virtually no spurious detections.

We believe the choice of the algorithm was not critical for the
dataset analized here. Due to the poor weather conditions during most
of the observing run and the resulting decreased photometric accuracy
on some nights and uneven temporal coverage, we did not expect to find
planetary transits. 

\begin{figure*}[hp]
\epsscale{2}
\plotone{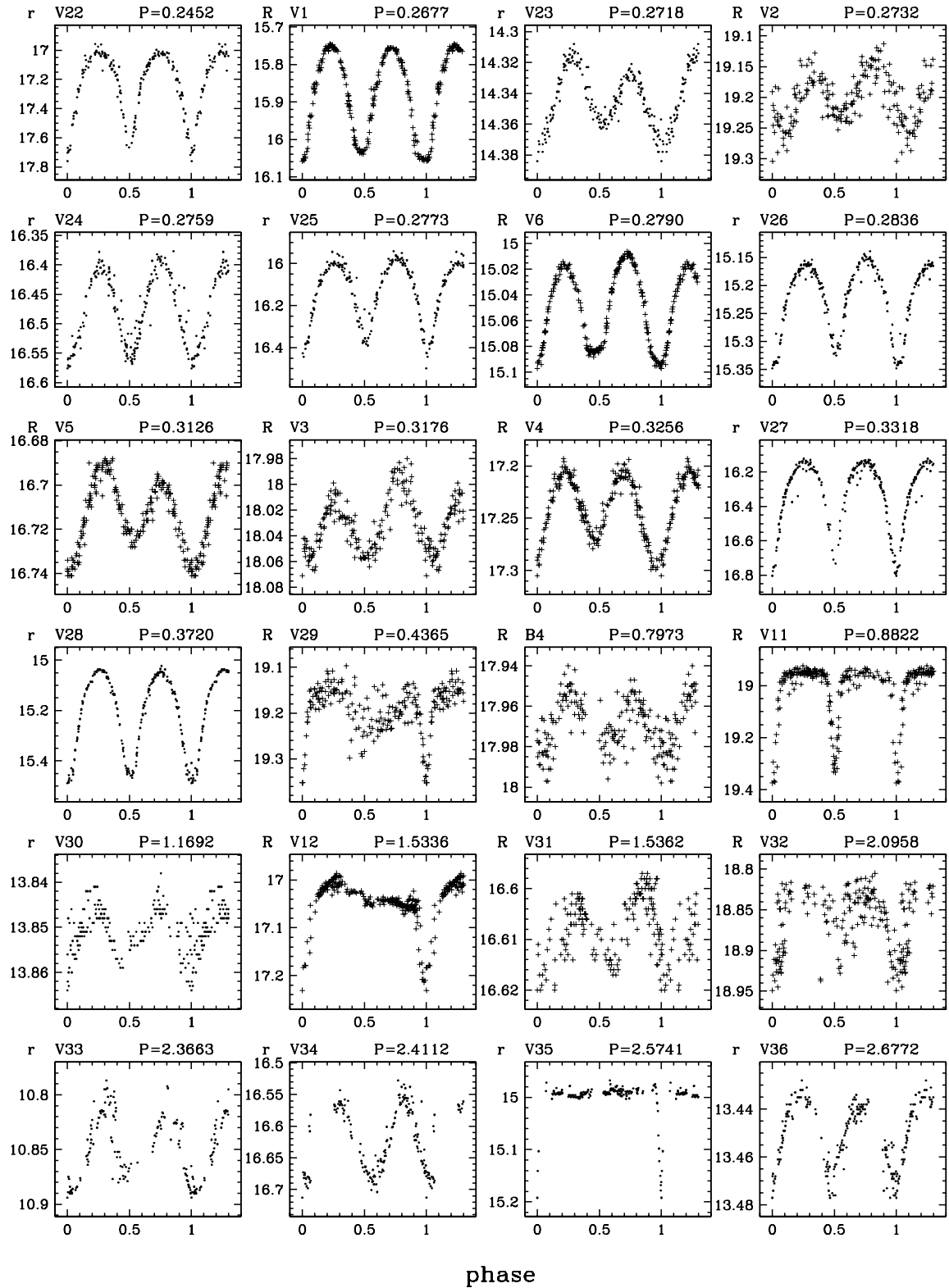}
\caption{The light curves of the 30 eclipsing binaries. Crosses
indicate standard $R$ magnitude, points - instrumental.}
\label{lc:ecl}
\end{figure*}

\addtocounter{figure}{-1}
\begin{figure*}[ht]
\plotone{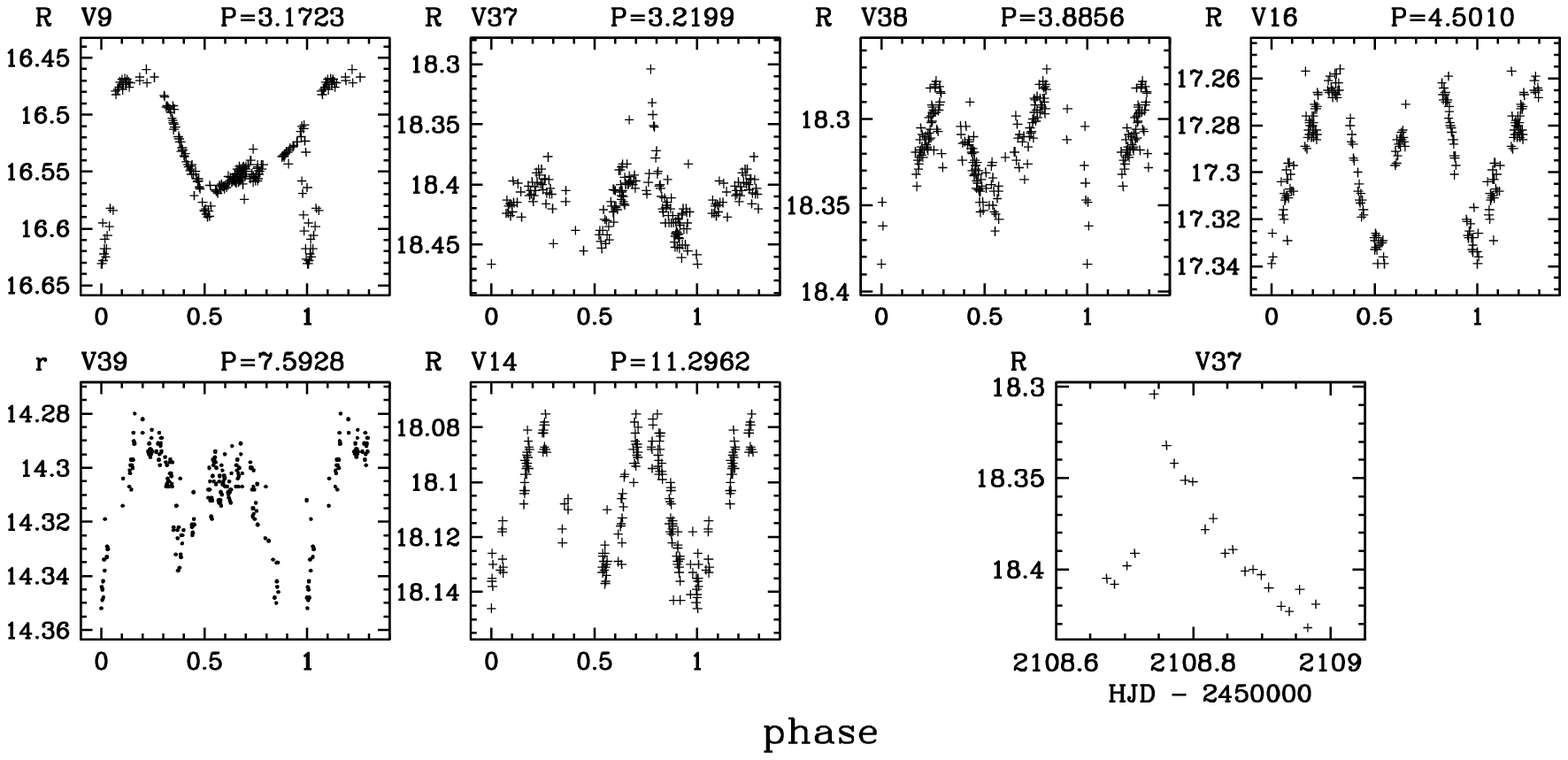}
\caption{Continued.}
\end{figure*}

\subsection{{\it Period Determination}}
To search for periodicities we have used the method introduced by
Schwarzenberg-Czerny (1996), employing periodic orthogonal polynomials
to fit the observations and the analysis of variance statistic to
evaluate the quality of the fit.

If observations $X$ consist of the sum of signal $F$ and noise $E$:
$X_k=F_k+E_k$, then the analysis of variance statistic
$\Theta(\omega)$ is defined as:
\[\Theta(\omega) \equiv \frac{\hat{Var}\{F\}}{\hat{Var}\{E\}} =
\frac{(K-2N-1)\sum_{n=0}^{2N}|c_n^2|}{2N[(X,X)-\sum_{n=0}^{2N}|c_n^2|]}\]
where 2N is the order of the complex polynomial (corresponding to a
Fourier series of N harmonics), K is the number of observations and
$c_n$ are the coefficients of the orthonormal polynomial
$\Psi_N(z)=\sum_{n=0}^{N} c_n \Phi_n(z)$, where
$\Phi_N(z)=\sum_{n=0}^{N} a_n^{(N)} z^n$ is the base. For the details
of the method the reader is referred to the Schwarzenberg-Czerny
(1996) paper.

\section{{\sc Variable Star catalog}}
\begin{figure*}[ph]
\plotone{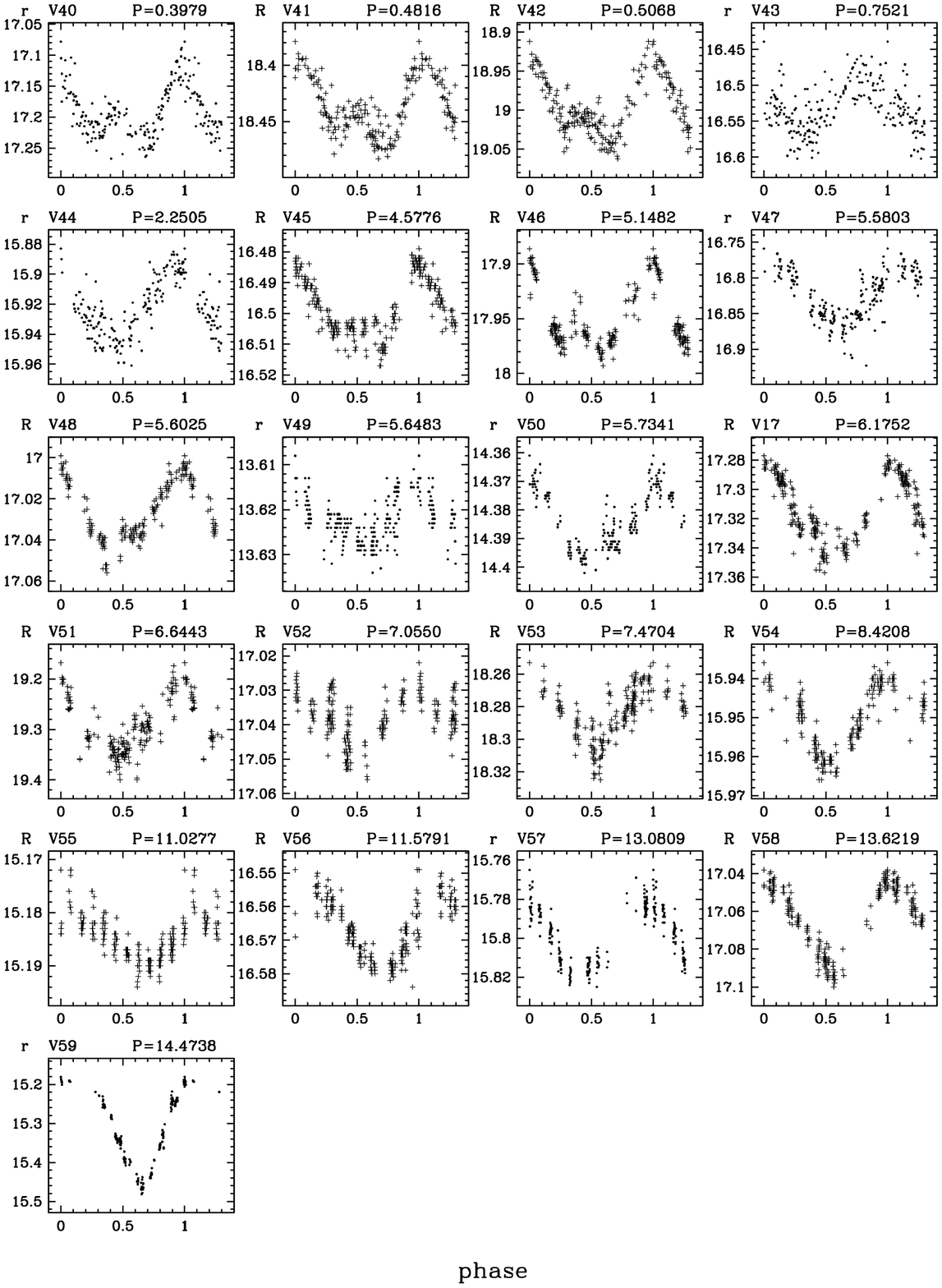}
\caption{The light curves of the 21 other periodic variables. Crosses
indicate standard $R$ magnitude, points - instrumental.}
\label{lc:pul}
\end{figure*}

We have found 62 variables in all four chips, 47 of them new ones (24
in chip 3): 30 eclipsing binaries, 21 other periodic variables and 11
miscellaneous ones. Their $R$-band light curves are shown in Figures
\ref{lc:ecl}, \ref{lc:pul} and \ref{lc:misc} and parameters listed in
Tables \ref{tab:ecl}, \ref{tab:pul} and \ref{tab:misc},
respectively.\footnote{The $R$ band photometry and finding charts
for all variables are available from the authors via the anonymous ftp
on cfa-ftp.harvard.edu, in the /pub/bmochejs/PISCES directory.}
The variables are also plotted on CMDs for their
respective chips [Figs \ref{cmd3} (chip 3) and \ref{cmd124} (chips 1,
2, 4)]. Eclipsing binaries are marked with circles, other periodic
variables with squares and miscellaneous variables with triangles.

We have recovered 18 of the 22 previously known variables. The missing
four are V7 and V8, which are outside of our feld of view, and V13 and
V19, which are saturated in the $R$-band data. The light curves of the
long period detached eclipsing binaries V10, V18 and V21 are also not
shown, as we have not observed any eclipses for them.

We have reobserved the W UMa type eclipsing variables V1-V6. We
confirm the longer of the two possible periods found for V5 by
RKH. This variable shows unequal maxima, as noted previously by
RKH. The shape of the light curve is seen to have changed: in our data
the variable attains higher maximum brightness at phase 0.25 (between
the primary and the secondary eclipse) and in the RKH data at phase
0.75. The light curve of V4 shows a similar change: the maxima are
almost equal now, while in the RKH data they were clearly uneven. The
variability of the maxima is probably due to changing starspots on the
surface of one or both of the components, like in RS CVn type
binaries. We have discovered six new W UMa type systems in chips 1, 2
and 4, V22-V28. One of them, V23, also shows unequal maxima.

V6 exhibits extremely small amplitudes of 0.1 mag in $R$ and $V$,
considering its flat-bottomed minima. Such light curve shape would
indicate an inclination angle close to $90^\circ$ and an extremely
small mass ratio for the binary components ($q<0.05$, Rucinski 1993).
A comparison with the 0.15 mag amplitude light curve in Kaluzny et
al. (1993) strongly suggests that this variable is blended in our
data. This is confirmed by the elongated appearance of the star on
our reference frame.

Another short period eclipsing variable, V29, does not seem to be a
contact system judging by the narrowness of the primary eclipse. On
the CMD it is located below the main sequence, together with V2 and
V11. It is most likely a field star behind the cluster. 

B4, one of the blue stars identified by KU92, shows brightness
variations with an amplitude of 0.05 mag. We have phased the light
curve with a period of 0.8 days, so that it displays two minima and
two maxima. B4 might be a non-eclipsing binary with brightness
modulations due to ellipsoidal variations or a reflection effect. An
argument in favor of this interpretation comes from a study of the
radial velocities of a sample of 70 sdBs which indicates that 45\% of
them are post-common envelope binaries with periods of the order of a
few hours to a few days (Saffer, Green \& Bowers 2000). The period for
B4 is consistent with this period range.

The binary nature of B4 is interesting from the point of view of the
origin of hot subdwarfs in metal rich clusters. Binary evolution has
been suggested as one of the possible formation mechanisms of such
objects (Moehler 2001).

\begin{figure*}[ph]
\plotone{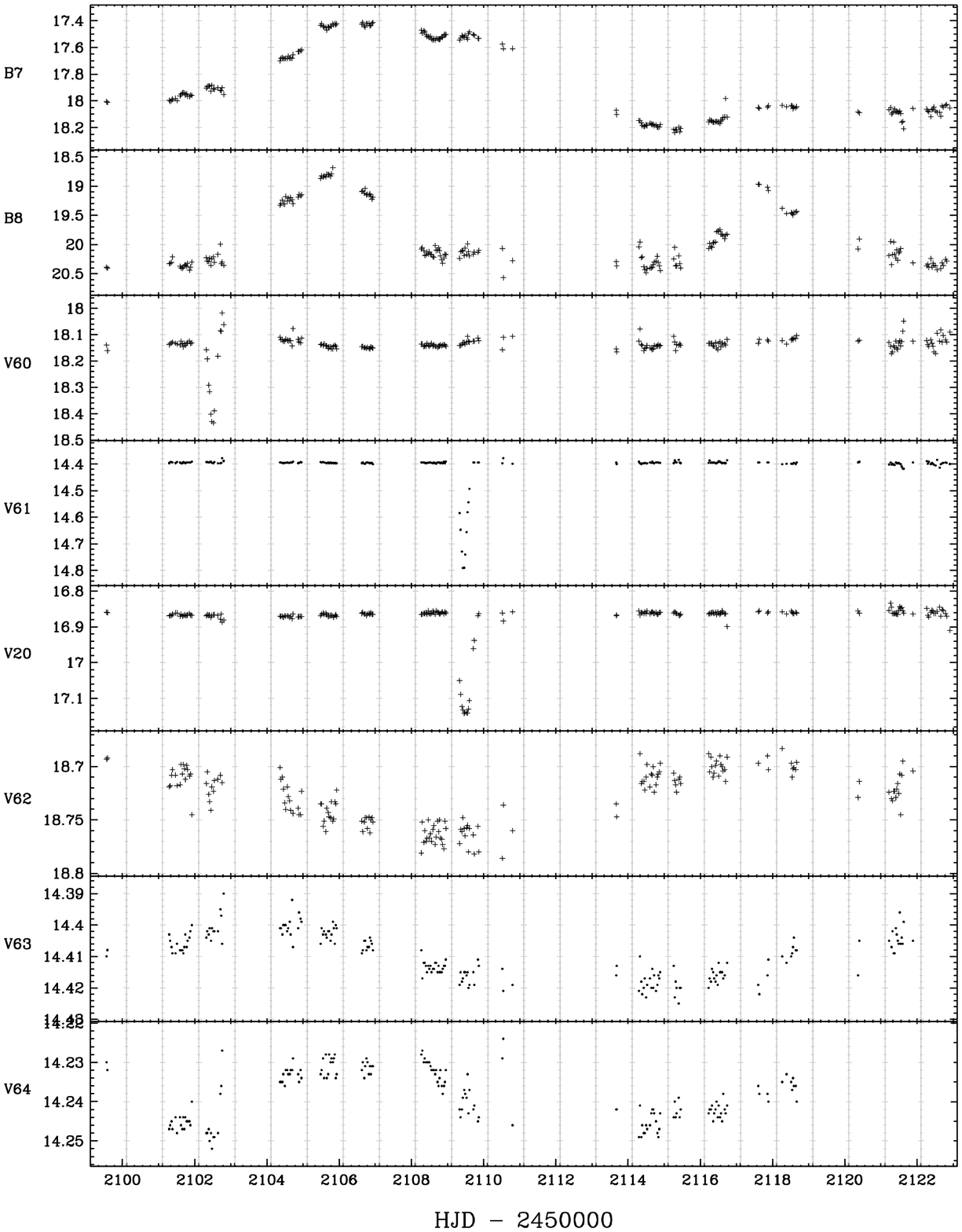}
\caption{The light curves of the 11 miscellaneous variables. Crosses
indicate standard $R$ magnitude, points - instrumental.}
\label{lc:misc}
\end{figure*}

\addtocounter{figure}{-1}
\begin{figure*}[th]
\plotone{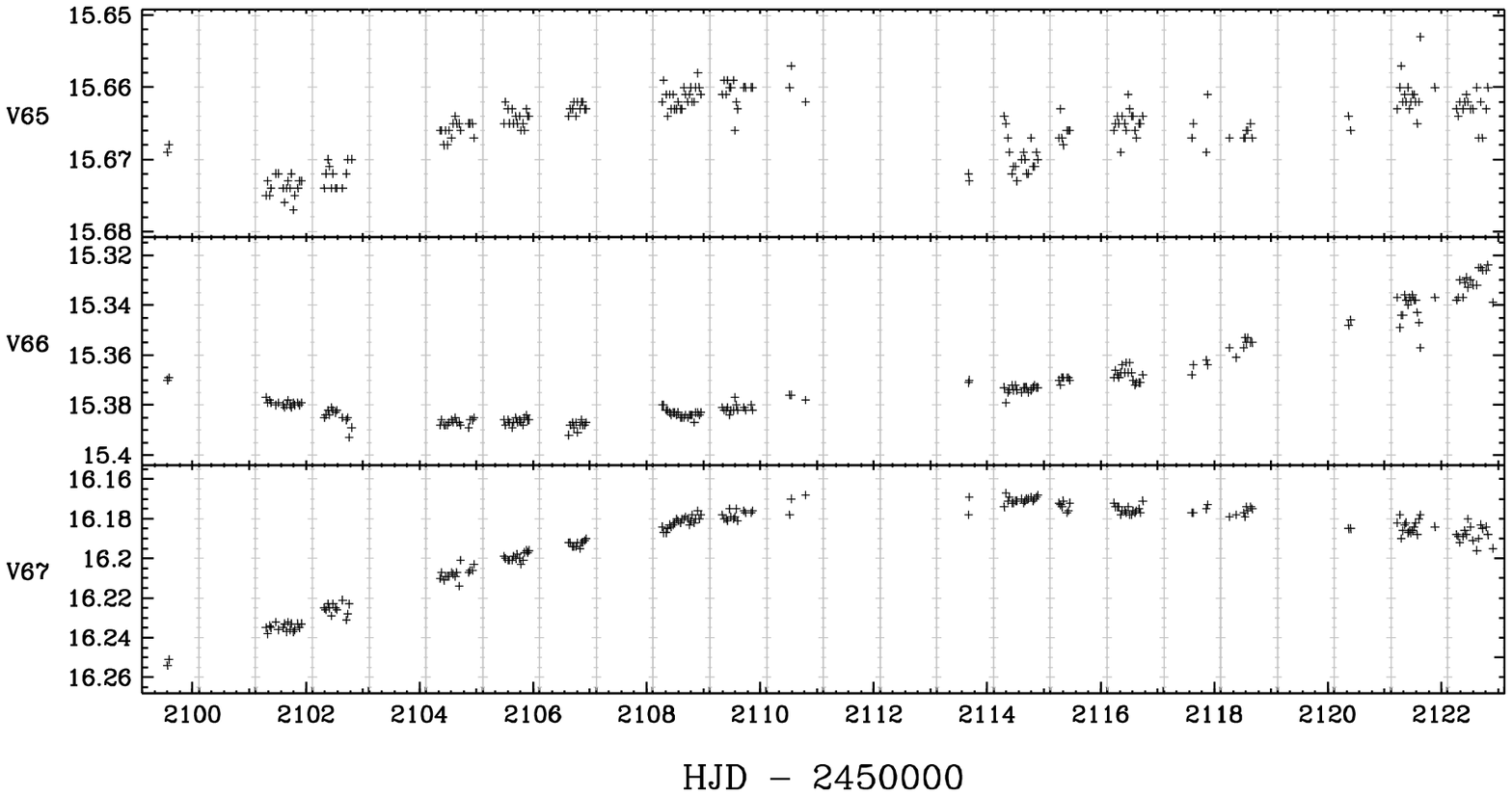}
\caption{Continued.}
\end{figure*}

We have also found some variables which could either be EB type
eclipsing binaries or ellipsoidal variables, as most have amplitudes
smaller than 0.1 mag. Among them are V14 and V16 with newly derived
periods. The new variables which belong to this type are V30, V31,
V33, V34, V36 and possibly V37 and V38. During one of the nights V37
seems to have undergone a UV Ceti-like outburst with an amplitude of
0.1 mag (Fig \ref{lc:ecl}). This variable is also very red, with
\vr=1.1 and most likely not a cluster member.

We have reobserved the RS CVn type eclipsing binary V9. The shape of
the S-wave does not seem to have changed significantly from the one
observed by RKH in 1996.

We have derived periods for the EA type eclipsing binaries V11, V12.
For the detached eclipsing system V20 we have observed only one eclipse.
Three new such sytems, V35, V60 and V61, were observed as well. A
period for V35 was derived, as it appears that one of the eclipses was
observed twice. Other possible EA type binaries are V32 and V39.

In addition to eclipsing binaries and ellipsoidal variables we have
also discovered periodic variables which seem to be of different
nature. The newly phased variable V17, as well as the new variables
V40-V44, V46, V47, V51, V53 are most likely BY Dra type variables --
rotating spotted late type dwarfs. On the CMDs they are located in the
vicinity of the lower MS ($V>18.5$) or redward of it. Some are
probably not cluster members. V40-V42 and possibly V46 exhibit small
humps between the maxima. Similar behavior has been observed for other
BY Dra type variables, eg. CC Eri (Amado et al. 2000). We have also
considered the possibility that V40-V42 are field RR Lyrae stars.
Despite the similarity in the light curve shape and period, this
interpretation seems unlikely because of their small amplitudes.

The remaining periodic variables are brighter than $V=18$. The nature
of their variability is unclear. These variables are spread throughout
the CMDs. V48, V52 and V58 seem to be located at the cluster turnoff,
and V45, V54, V56, V49 and V50 above it. V55, significantly redder
than the RGB stars, is probably not a cluster member. The variables
V57 and V59 seem to be located redward of the turnoff. V59 exibits
quite a high amplitude of 0.3 mag. 

Among the miscellaneous variables there are two especially interesting
ones: the cataclysmic variables B7 (V15) and B8 (Kaluzny et
al. 1997). B7 was observed to undergo a 0.6 mag outburst while in its
high state. B8 showed two dwarf-novae type outbursts, the first one
with an amplitude of 2 mag and duration of about 5 days and a second,
slightly fainter one. 

Five of the six other miscellaneous variables, V63-V67, are located
above the cluster turnoff. V62 appears to be located on the MS. Most
of the bright variables seem to show long period or quasi-periodic
variability. V66 and V67, located $\sim 0.1$ mag redward of the RGB,
display variations on a longer timescale than the other variables.

\section{{\sc Conclusions}}
In this paper we have demonstrated the feasibility of obtaining
photometry accurate enough to detect planets through transits in open
clusters with 1 m class telescopes. The analysis of the data collected
for this purpose resulted in the discovery of 47 new low amplitude
variables, compared to 22 previously known (Kaluzny \& Rucinski 1993,
RKH).

The first stage of our project will be to obtain for this and several
other clusters continuous observations with a 1-m class telescope for
about $\sim$30 nights under very good observing conditions. The
photometry will be obtained in two bands, $R$ and $V$, to allow us to
differentiate between planetary transits and blended eclipsing
binaries. Planetary transits should be gray, as the planet does not
contribute any measurable light to the system, while the superposition
of an eclipsing binary with another star will show a change in color
during the eclipse.

In the next stage better light curves will be obtained with a 2 m
class telescope for selected planet transit candidates. Using radial
velocity measurements derived from spectroscopic observations with a
6-10 m class telescope it will be possible to distinguish planetary
and brown dwarf transits from grazing eclipses by main sequence
companions. A precision of 1 km s$^{-1}$ will enable us to identify
and reject stars with companions above 0.0075 M$_\odot$ ($\sim 8$
M$_{J}$).

\acknowledgments{ We would like to thank Janusz Kaluzny for helpful
discussions and careful reading of the manuscript and Alceste Bonanos
for her help in obtaining some of the data. We also thank the referee,
Ronald Gilliland, for a very useful report which significantly
improved the paper. Part of this work was done during BJM's visit to
the Harvard-Smithsonian Center for Astrophysics. BJM was supported by
the Polish KBN grant 5P03D004.21 and the Foundation for Polish Science
stipend for young scientists. DDS acknowledges support from the Alfred
P. Sloan Foundation and from NSF grant No. AST-9970812.}

\begin{deluxetable}{rrrrlccc}
\tabletypesize{\footnotesize}
\tablewidth{25pc}
%\tablewidth{28.5pc}
\tablecaption{Eclipsing binaries in NGC 6791}
\tablehead{\colhead{ID} & \colhead{$\alpha_{2000}$} &
\colhead{$\delta_{2000}$} &\colhead{P [d]} & \colhead{$R_{max}$} &
\colhead{$V_{max}$} &\colhead{$A_R$} & \colhead{$A_V$}}
\startdata
 V22 & 19.338527 & 37.508317 &  0.2452 & 16.507$^*$ & 19.239  &  0.794 &  0.862  \\
  V1 & 19.346557 & 37.742245 &  0.2677 & 15.744     & 16.223  &  0.327 &  0.328  \\
 V23 & 19.338612 & 37.787813 &  0.2718 & 13.772$^*$ & 16.255  &  0.072 &  0.064  \\
  V2 & 19.354874 & 37.766766 &  0.2732 & 19.125     & 19.623  &  0.176 &  0.281  \\
 V24 & 19.332915 & 37.595612 &  0.2759 & 15.897$^*$ & 18.546  &  0.200 &  0.209  \\
 V25 & 19.328430 & 37.713417 &  0.2773 & 15.414$^*$ & 18.004  &  0.540 &  0.525  \\
  V6 & 19.350753 & 37.813595 &  0.2790 & 15.008     & 15.397  &  0.093 &  0.097  \\
 V26 & 19.345806 & 37.561879 &  0.2836 & 14.626$^*$ & 16.659  &  0.210 &  0.221  \\
  V5 & 19.346259 & 37.813295 &  0.3126 & 16.687     & 17.149  &  0.055 &  0.060  \\
  V3 & 19.354378 & 37.769396 &  0.3176 & 17.986     & 18.550  &  0.087 &  0.128  \\
  V4 & 19.348395 & 37.806623 &  0.3256 & 17.194     & 17.710  &  0.109 &  0.088  \\
 V27 & 19.336295 & 37.649081 &  0.3318 & 15.624$^*$ & 18.082  &  0.698 &  0.651  \\
 V28 & 19.328836 & 37.591770 &  0.3720 & 14.540$^*$ & 16.949  &  0.474 &  0.454  \\
 V29 & 19.354795 & 37.751474 &  0.4365 & 19.117     & 19.653  &  0.248 &  0.359  \\
  B4 & 19.353583 & 37.764289 &  0.7973 & 17.945     & 17.797  &  0.054 &  0.064  \\
 V11 & 19.342582 & 37.804615 &  0.8822 & 18.911     & 19.424  &  0.462 &  0.395  \\
 V30 & 19.328617 & 37.501978 &  1.1692 & 13.359$^*$ & 15.554  &  0.023 &  0.030  \\
 V12 & 19.345259 & 37.849032 &  1.5336 & 16.989     & 17.508  &  0.241 &  0.238  \\
 V31 & 19.350685 & 37.785921 &  1.5362 & 16.597     & 17.067  &  0.024 &  0.040  \\
 V32 & 19.341003 & 37.787297 &  2.0958 & 18.809     & 19.242  &  0.140 &  0.174  \\
 V33 & 19.344394 & 37.731803 &  2.3663 & 10.796$^*$ & \nodata &  0.095 & \nodata \\
 V34 & 19.335880 & 37.736360 &  2.4112 & 15.993$^*$ & 18.597  &  0.175 &  0.197  \\
 V35 & 19.345574 & 37.511924 &  2.5741 & 14.454$^*$ & 16.435  &  0.220 &  0.219  \\
 V36 & 19.332336 & 37.570224 &  2.6772 & 12.950$^*$ & 16.138  &  0.047 &  0.092  \\
  V9 & 19.346634 & 37.777067 &  3.1723 & 16.463     & 17.146  &  0.173 &  0.154  \\
 V37 & 19.355069 & 37.851982 &  3.2199 & 18.385     & 19.436  &  0.080 &  0.291  \\
 V38 & 19.351024 & 37.768325 &  3.8856 & 18.276     & \nodata &  0.097 & \nodata \\
 V16 & 19.352108 & 37.802683 &  4.5010 & 17.258     & 17.749  &  0.085 &  0.097  \\
 V39 & 19.350133 & 37.639703 &  7.5928 & 13.764$^*$ & 15.964  &  0.068 &  0.053  \\
 V14 & 19.347684 & 37.756898 & 11.2962 & 18.075     & 18.615  &  0.071 &  0.107  \\
\enddata
\tablecomments{$^*$ Instrumental magnitudes}
\label{tab:ecl}
\end{deluxetable}

\begin{deluxetable}{rrrrlccc}
\tabletypesize{\footnotesize}
\tablewidth{25pc}
%\tablewidth{28.5pc}
\tablecaption{Other periodic variables in NGC 6791}
\tablehead{\colhead{ID} & \colhead{$\alpha_{2000}$} &
\colhead{$\delta_{2000}$} &\colhead{P [d]} & \colhead{$R_{max}$} &
\colhead{$V_{max}$} &\colhead{$A_R$} & \colhead{$A_V$}}
\startdata
 V40 & 19.327499 & 37.617002 &  0.3979 & 16.714$^*$ & 19.340  &  0.060 &  0.070  \\
 V41 & 19.347492 & 37.806861 &  0.4816 & 18.436     & 19.201  &  0.040 &  0.040  \\
 V42 & 19.350057 & 37.714872 &  0.5068 & 18.993     & 19.641  &  0.060 &  0.070  \\
 V43 & 19.344330 & 37.641893 &  0.7521 & 16.003$^*$ & 19.026  &  0.040 &  0.100  \\
 V44 & 19.326974 & 37.694940 &  2.2505 & 15.383$^*$ & 17.809  &  0.030 &  0.030  \\
 V45 & 19.346126 & 37.701676 &  4.5776 & 16.499     & 16.982  &  0.010 &  0.020  \\
 V46 & 19.355274 & 37.798960 &  5.1482 & 17.947     & 18.642  &  0.040 &  0.050  \\
 V47 & 19.327533 & 37.536375 &  5.5803 & 16.345$^*$ & 19.378  &  0.050 &  0.070  \\
 V48 & 19.352077 & 37.718522 &  5.6025 & 17.028     & 17.477  &  0.020 &  0.020  \\
 V49 & 19.341914 & 37.614271 &  5.6483 & 13.102$^*$ & 15.114  &  0.010 &  0.010  \\
 V50 & 19.343115 & 37.517920 &  5.7341 & 13.865$^*$ & 15.836  &  0.010 &  0.020  \\
 V17 & 19.344137 & 37.817929 &  6.1752 & 17.317     & 17.958  &  0.030 &  0.040  \\
 V51 & 19.353380 & 37.748568 &  6.6443 & 19.294     & 19.892  &  0.090 &  0.100  \\
 V52 & 19.355798 & 37.772043 &  7.0550 & 17.039     & 17.465  &  0.010 &  0.010  \\
 V53 & 19.350232 & 37.743195 &  7.4704 & 18.283     & 18.720  &  0.020 &  0.020  \\
 V54 & 19.355196 & 37.726814 &  8.4208 & 15.950     & 16.460  &  0.010 &  0.010  \\
 V55 & 19.356229 & 37.841641 & 11.0277 & 15.185     & 16.120  &  0.000 &  0.010  \\
 V56 & 19.345907 & 37.763584 & 11.5791 & 16.567     & 17.034  &  0.010 &  0.020  \\
 V57 & 19.349417 & 37.518655 & 13.0809 & 15.278$^*$ & 17.689  &  0.020 &  0.030  \\
 V58 & 19.354038 & 37.801263 & 13.6219 & 17.069     & 17.498  &  0.030 &  0.030  \\
 V59 & 19.339303 & 37.806090 & 14.4738 & 14.759$^*$ & 17.385  &  0.150 &  0.170  \\
\enddata
\tablecomments{$^*$ Instrumental magnitudes}
\label{tab:pul}
\end{deluxetable}

\begin{deluxetable}{rrrlccc}
\tabletypesize{\footnotesize}
\tablewidth{22pc}
%\tablewidth{25pc}
\tablecaption{Other periodic variables in NGC 6791}
\tablehead{\colhead{ID} & \colhead{$\alpha_{2000}$} &
\colhead{$\delta_{2000}$} & \colhead{$R_{max}$} &
\colhead{$V_{max}$} &\colhead{$A_R$} &\colhead{$A_V$}}
\startdata
 V61 & 19.328572 & 37.485425 & 13.903$^*$ & 16.383  &  0.470 &  0.383  \\
 V63 & 19.327780 & 37.495899 & 13.913$^*$ & 16.603  &  0.029 &  0.035  \\
 V64 & 19.353181 & 37.498769 & 13.707$^*$ & 15.773  &  0.024 &  0.085  \\
 V62 & 19.350848 & 37.731070 & 18.686     & 19.215  &  0.098 &  0.095  \\
  B8 & 19.343257 & 37.747859 & 18.803     & 19.117  &  1.951 &  2.991  \\
 V66 & 19.352340 & 37.748699 & 15.324     & 16.056  &  0.067 &  0.034  \\
 V20 & 19.348416 & 37.759650 & 16.846     & 17.327  &  0.314 &  0.293  \\
 V60 & 19.350193 & 37.762524 & 18.067     & 18.633  &  0.346 &  0.217  \\
 V65 & 19.347908 & 37.791815 & 15.658     & 16.211  &  0.018 &  0.025  \\
  B7 & 19.352056 & 37.799037 & 17.418     & 17.476  &  0.880 &  0.883  \\
 V67 & 19.351021 & 37.801035 & 16.167     & 16.845  &  0.079 &  0.096  \\
\enddata
\tablecomments{$^*$ Instrumental magnitudes}
\label{tab:misc}
\end{deluxetable}

\end{document}